\documentclass[useAMS,usenatbib]{mn2e}

\DeclareSymbolFont{cmletters}{OML}{cmm}{m}{it}
\DeclareMathSymbol{v}{\mathalpha}{cmletters}{"76}

\usepackage{tabularx,ragged2e,booktabs,caption}
\usepackage{amsmath}
\usepackage{amssymb}
\usepackage{epsfig}
\usepackage{graphicx}
\usepackage{ifthen}
\usepackage{latexsym}
\usepackage{rotating}
\usepackage{subfigure}
\usepackage{times,epsf}
\usepackage{txfonts}
\usepackage{varioref}
\usepackage{verbatim}
\usepackage{array}
\usepackage{url}
\usepackage{color}
\usepackage[dvipsnames]{xcolor}
\usepackage[T1]{fontenc}

\newcommand{\be}{\begin{equation}}
\newcommand{\ee}{\end{equation}}
\newcommand{\bea}{\begin{eqnarray}}
\newcommand{\eea}{\end{eqnarray}}

\newcommand\apj{Astrophysical Journal}
\newcommand\apjl{Astrophysical Journal Letters}

\newcommand\apjs{Astrophysical Journal Suppl. Ser.}

\newcommand\aap{Astronomy \& Astrophysics}

\newcommand\mnras{Monthly Notices of the Royal Astronomical Society}

\newcommand\pasj{Publications of the Astronomical Society of Japan}

\newcommand\ARAA{Ann. Rev. Astron. Asrophys.}
\newcommand\jqsrt{Journal of Quantitative Spectroscopy and Radiative Transfer}
\newcommand\araa{\ARAA}

\newcommand{\Medd}{\dot M_{\rm Edd}}

\title[Comptonization in accretion disks]{Photon-conserving
  Comptonization in simulations of accretion disks around black holes}
\author[A. S\k{a}dowski, R. Narayan]
       {Aleksander S\k{a}dowski$^1$\footnotemark[1]	        
        and Ramesh Narayan$^{1}$\thanks{E-mail: asadowsk@mit.edu (AS); 
rnarayan@cfa.harvard.edu (RN);} \\
        $^1$ MIT Kavli Institute for Astrophysics and Space Research,
77 Massachusetts Ave, Cambridge, MA 02139, USA\\
$^2$ Harvard-Smithsonian Center for Astrophysics, 60 Garden St., Cambridge, MA 02138, USA}

\begin{document}

\maketitle

\label{firstpage}

\begin{abstract}

We introduce a new method for treating Comptonization in computational
fluid dynamics. By construction, this method conserves the number of
photons. Whereas the traditional ``blackbody Comptonization'' approach
assumes that the radiation is locally a perfect blackbody and
therefore uses a single parameter, the radiation temperature, to
describe the radiation, the new ``photon-conserving Comptonization''
approach treats the photon gas as a Bose-Einstein fluid and keeps
track of both the radiation temperature and the photon number
density. We have implemented photon-conserving Comptonization in the
general relativistic radiation magnetohydrodynamical code
\texttt{KORAL} and we describe its impact on simulations of mildly
super-critical black hole accretion disks. We find that blackbody
Comptonization underestimates the gas and radiation temperature by up
to a factor of two compared to photon-conserving Comptonization. This
discrepancy could be serious when computing spectra. The
photon-conserving simulation indicates that the spectral color
correction factor of the escaping radiation in the funnel region
of the disk could be as large as 5.
\end{abstract}


\section{Introduction}
\label{s.introduction}

Accretion on black holes (BHs) is an important energy source that
shapes the evolution of the Universe. Accreting supermassive BHs in the
centers of galaxies expel a significant fraction of their accreted rest
mass energy in the form of kinetic, thermal and radiative feedback, which
affects the evolution of their host galaxies \citep[e.g.,][]{fabian-12,king+15}. At the same time, stellar mass BHs
accreting in compact binaries, although not powerful enough to influence their
hosts, are often the brightest X-ray sources in their galaxies. 

For long our understanding of accretion was based on
analytical models, most importantly the thin disk model of
\cite{ss73} and \cite{novikovthorne73}. Although such models provide important information on the dynamical properties of the accretion disk and
predict its radiative luminosity, they provide only a rough idea of
the emitted electromagnetic spectrum. To calculate the spectrum properly,
including the effect of spectral hardening due to Comptonization in the strongly scattering medium, one
has to solve the full radiative transfer problem for the disk atmosphere. Valuable
work has been done in this area by \cite{davis+05}, who computed the
  vertical structure of disk atmospheres with a self-consistent
  treatment of radiative transfer. Their work made use of the code TLUSTY
  \citep{hubenylanz-95}, with modifications to treat relativistic disks as described in
  \cite{hubeny2-97} and \cite{hubeny+01}. Spectra obtained with this approach have been successfully
used for modelling disk emission in the thermal-dominant (high/soft) state of X-ray binaries,
and in particular for estimating BH spins via the continuum fitting
method \citep[e.g.,][]{mcclintock+spins}.

The thin disk model can be applied only to optically thick,
geometrically thin accretion flows. Once the accretion rate
exceeds the Eddington rate, the various assumptions behind the thin disk model
(e.g., radiative efficiency, Keplerian rotation) break down. Super-critical 
accretion flows have been modeled semi-analytically
\citep{abra88,sadowski.phd}, but these models do not consistently account
for mass outflows. In addition, although
super-critical disks are known to be geometrically thick, their spectra have in the past been
calculated using the plane-parallel approximation \citep[see][]{straub+11},
which is a gross simplifcation.

The proper, although computationally expensive, approach to
modeling accretion disks is to simulate
the accretion flow in multiple dimension, including angular momentum
transport via
magnetically-driven turbulence \citep{balbushawley-98}, and to solve in parallel the
multi-dimensional radiative transfer problem. Numerical codes to carry out
such computations have become
available only recently. Because of computational cost, often the radiative
transfer in such simulations is simplified, and only in the
postprocessing stage is the full, frequency dependent radiative tranport
computed.
The first magnetohydrodynamical numerical simulations of optically thick accretion disks,
including radiation, were performed by \cite{ohsuga09} and \cite{ohsuga11}. These authors
used a Newtonian approximation for gravity and modeled radiation under the flux limited
diffusion approximation. In recent
years, other groups have developed methods for general relativistic
radiation-MHD simulations, including better radiation closures
\citep{sadowski+koral, mckinney+harmrad, fragile+14}. Newtonian simulations
with advanced
radiative transport methods have also been carried out \citep{jiang+14a}.

Super-critical accretion flows around stellar-mass BHs are very optically thick to electron
scattering. They have temperatures of order a keV or more, so their
absorption opacity is subdominant to scattering. As a result one expects significant
Comptonization and spectral hardening of the escaping radiation. So far,
however, simulations of super-critical disks have either ignored
Comptonization altogether
\citep{sadowski+koral2,jiang+14b} or have treated it in
a crude manner, which we call ``blackbody Comptonization'', in which
the radiation is assumed to be locally
a perfect blackbody
\citep{kawashima+09, sadowski+dynamo}.

It might be hoped that it is sufficient to carry out simulations
with the above ``blackbody Comptonization,'' followed by a more detailed
proper Comptonization calculation in a later
post-processing stage. However, as we show in this paper,
the thermal
state of the gas depends strongly on the particular
implementation of Comptonization. The blackbody Comptonization 
approximation should therefore be treated with caution.

In this work we introduce an improved model of Comptonization in
which we avoid the assumption that the local radiation at each point has a
perfect blackbody spectrum. In the new method, we independently evolve both the
radiation energy density and the photon number density, thereby relaxing the
previous assumption that the two quantities are related precisely by the
Planck function.  To assess what effect this improvement in the modeling of
Comptonization has, we have performed three simulations of a
super-critical accretion disk, one with our previous ``blackbody Comptonization,'' one with the improved ``photon-conserving
Comptonization'' described in this paper, and one with no Comptonization at all. We compare the
properties of the three simulations and draw conclusions.

The paper is organized as follows. In Section~\ref{s.comptonization} we
introduce the blackbody and photon-conserving versions of Comptonization.  In
Section~\ref{s.method} we give implementation details of the
photon-conserving approach. In Section~\ref{s.sims} we present
the three simulations and analyze the results, and in Section~\ref{s.discussion} 
we briefly discuss the implications of the results.

\section{Comptonization}
\label{s.comptonization}

\subsection{Basic equations of GRRMHD}

Within the framework of general relativity, the interaction between
gas and radiation is described through the following general
relativistic radiation magnetohydrodynamics (GRRMHD) conservation
equations for gas density, $\rho$, and the
stress-energy tensors, $T_\nu^\mu$ and $R_\nu^\mu$, of the magnetized
gas and radiation \citep{sadowski+koral,sadowski+koral2},
\begin{eqnarray}
\hspace{2.cm}\left(\rho u^\mu\right)_{;\mu} &=& 0, \label{rhocons} \\
\left(T_\nu^\mu\right)_{;\mu} &=& G_\nu, \label{Tmunu} \\
\left(R_\nu^\mu\right)_{;\mu} &=& -G_\nu. \label{Rmunu}
\end{eqnarray}
Here $G^\nu$ is the radiation four-force density acting on the gas. In
orthonormal coordinates in the fluid (or gas) frame (denoted by hats), and in the absence of
Comptonization, we have
\begin{eqnarray}
\hspace{2.cm}\widehat{G}^0 &=& \kappa_{\rm a}\rho (\widehat{E}-4\pi \widehat{B}), \label{G00}
\\ \widehat{G}^i &=& (\kappa_{\rm a}+\kappa_{\rm es}) \rho \widehat{F}^i.
\label{Gi}
\end{eqnarray}
Here $\kappa_{\rm a}$ is the Rosseland mean absorption opacity
  (cross-sections per unit mass), which we adopt for simplicity instead
of the proper absorption and Planck mean opacities. $\kappa_{\rm es}$ is the scattering opacity, $\widehat E$ is
the radiative energy density in the fluid frame,  $\widehat B$
is the intensity of 
blackbody radiation for gas of temperature $T_{\rm g}$, 
\be
\label{Tg}
\widehat B=\frac {a
T_{\rm g}^4}{4\pi},
\ee
where $a$ is the radiation constant,
and $\widehat F^i$
is the radiative flux three-vector.

Equation (\ref{G00}) describes the rate of change of the fluid energy
density as a result of energy gain through absorption, $\kappa_{\rm
  a}\rho\widehat{E}$, and energy loss through emission, $\kappa_{\rm
  a}\rho (4\pi \widehat{B})$.
The simplest approximation for modeling radiation is to treat it as a local 
blackbody at each point, so that the energy distribution is fully
described by the local radiation temperature $T_{\rm r}$.  In this
approximation, the radiation energy density is given by
\begin{equation}
\widehat{E} = aT_{\rm r}^4. \label{Tr1}
\end{equation}
Correspondingly, we rewrite equation (\ref{G00}) as
\begin{equation}
\widehat{G}^0 = \kappa_{\rm a} \rho a (T_{\rm r}^4-
T_{\rm g}^4). \label{G0}
\end{equation}
Equation
(\ref{G0}) shows that gas gains energy at a rate proportional to
$T_{\rm r}^4$ and loses energy proportional to $T_{\rm g}^4$. As a
result, the two temperatures are pushed towards each other, i.e., the
system is driven towards thermal equilibrium. Note that $\kappa_{\rm
  es}$ does not appear in the energy equation. For the present discussion,
which ignores Comptonization, we have pure Thomson scattering, which only redirects
photons but does not change photon energies.

Equation (\ref{Gi}) describes the rate of change of the fluid momentum
density. The gas acquires the momentum of each photon that it either
absorbs or scatters. This explains the presence of both $\kappa_{\rm
  a}$ and $\kappa_{\rm es}$ in this equation. However, radiation that
is emitted or scattered by the gas is symmetric in the fluid
frame. Therefore, there is no corresponding momentum loss, 
hence no counter-balancing term with a negative sign.  Note
that the fluid gains momentum density in a direction parallel to the
radiation flux three-vector $\hat{F}^i$, and the radiation loses a corresponding
amount of momentum density. Hence the system is driven towards a state
in which there is no relative motion between the fluid and radiation
frames, i.e., no radiation flux in the fluid frame.

\subsection{Thermal Comptonization: Blackbody Approximation}
\label{s.thermal}

The main effect of Comptonization is that scattering causes not just
momentum transfer between the radiation and gas, but also energy
transfer. A soft photon of energy $\epsilon_0$ which scatters off a
thermal electron with temperature $T_{\rm e}$ on average gains an
energy $\langle\Delta\epsilon\rangle$ given by the following
expressions in the non-relativistic and ultra-relativistic limits,
respectively \citep{rybicki-book},
\begin{eqnarray}
\langle\Delta\epsilon\rangle &=& \left(\frac{4kT_{\rm e}}{m_{\rm e}c^2}\right)\epsilon_0,
\qquad 4kT_{\rm e} \ll m_{\rm e}c^2, \label{CompNR} \\
\langle\Delta\epsilon\rangle &=& \left(\frac{4kT_{\rm e}}{m_{\rm e}c^2}\right)^2\epsilon_0,
\qquad 4kT_{\rm e} \gg m_{\rm e}c^2. \label{CompR}
\end{eqnarray}
For a general temperature, the result is given in equation (2.43) in
\cite{pozdnyakov+83}. A good fitting
function (maximum fractional error 1.2\%) is the following:
\begin{equation}
\langle\Delta\epsilon\rangle = \epsilon_0 
\left(\frac{4kT_{\rm e}}{m_{\rm e}c^2}\right) 
\left[1+3.683 \left(\frac{kT_{\rm e}}{m_{\rm e}c^2}\right) 
+4 \left(\frac{kT_{\rm e}}{m_{\rm e}c^2}\right)^2\right]
\left[1+\left(\frac{kT_{\rm e}}{m_{\rm e}c^2}\right)\right]^{-1},
\label{Compgen} \\
\end{equation}

The above expressions are valid so long as the photon is soft, i.e.,
the radiation temperature is much less than the gas temperature. When
the two temperatures are equal, thermodynamics guarantees that there
is no energy transfer between gas and radiation. Similarly, when the
radiation temperature is larger than the gas temperature, we expect
energy to flow from the radiation to the gas. To allow for these
effects, we modify the expression for $\widehat{G}^0$ in equation
(\ref{G0}) to the following:
\begin{eqnarray}
\widehat{G}^0 &=& \rho a \kappa_{\rm a}(T_{\rm r}^4-
T_{\rm g}^4)
- \kappa_{\rm es}\rho\widehat{E} \left[\frac{4k(T_{\rm g}-T_{\rm
      r})}{m_{\rm g}c^2}\right]  \nonumber \\
&~~&\times\left[1+3.683 \left(\frac{kT_{\rm g}}{m_{\rm g}c^2}\right) 
+4 \left(\frac{kT_{\rm g}}{m_{\rm g}c^2}\right)^2\right]
\left[1+\left(\frac{kT_{\rm g}}{m_{\rm g}c^2}\right)\right]^{-1},
\label{CompG0}
\end{eqnarray}
where we have replaced $T_e$ by $T_{\rm g}$.
The negative sign in the Compton term is because gas cools when
$T_{\rm g} > T_{\rm r}$. The cooling is proportional to the radiation
energy density, $\widehat{E}$, and to the number of scatterings per
unit time, $c\kappa_{\rm es}\rho$. The final two factors in square
parentheses correspond to the approximate correction for relativistic
temperatures given in equation~(\ref{Compgen}). Apart from these factors, equation (\ref{CompG0}) is
identical to the prescription used by \cite{kawashima+09}.

As far as the momentum equation is concerned, we assume that the
Compton-scattered radiation is symmetric in the fluid frame and
carries no net momentum (a fairly good approximation in the soft
photon limit).  Under this approximation, we keep equation (\ref{Gi})
unchanged. 

Thus, in this (simplest) version of thermal Comptonization, which we call
``blackbody Comptonization,'' we solve
equations (\ref{Tmunu}) and (\ref{Rmunu}) using equations
(\ref{CompG0}) and (\ref{Gi}), where $T_{\rm g}$ and $T_{\rm r}$ are
given by equations~(\ref{Tg}) and (\ref{Tr1}), respectively. The main weakness of this approach is that it does not
conserve photon number during scattering. Instead it assumes that
the radiation is a perfect
blackbody and uses equation (\ref{Tr1}) to obtain the radiation
temperature.

\subsection{Photon-conserving Comptonization}
\label{s.ncompton}

In the next approximation, which we call ``photon-conserving Comptonization,'' we give up the assumption that the spectral
shape is a perfect blackbody, but instead assume that the photons have a
Bose-Einstein (BE) distribution in the fluid frame, described by two
parameters: radiation temperature $T_r$ and radiation chemical
potential $\mu$.  Correspondingly, we keep track of not only
the radiation energy and momentum density, $\widehat{E}$,
$\widehat{F}^i$, but also the photon number density,
$\widehat{n}$. This generalization allows us to include the effect of
radiation dilution and spectral hardening.

Let us define the dimensionless photon energy, $x=h\nu/kT_{\rm r}$, and the
dimensionless chemical potential, $\xi=\mu/kT_{\rm r}$. The
angle-integrated BE distribution $f(x)dx$ of photons at temperature
$T_{\rm r}$ is given by
\begin{equation}
f(x)dx = C(kT_{\rm r})^3 e^{-\xi} \frac{x^2dx}{e^x-e^{-\xi}}, \qquad C
\equiv \frac{8\pi}{c^3h^3}. \label{BE}
\end{equation}
The radiation energy density and photon number density are then
\begin{equation}
\widehat{E} = C (kT_{\rm r})^4 e^{-\xi} I_E(\xi), \qquad
I_E(\xi) = \int_0^\infty \frac{x^3dx}{e^x-e^{-\xi}},
\end{equation}
\begin{equation}
\widehat{n} = C (kT_{\rm r})^3 e^{-\xi} I_N(\xi), \qquad
I_N(\xi) = \int_0^\infty \frac{x^2dx}{e^x-e^{-\xi}}.
\end{equation}
In the Planck limit, i.e., when $\xi=0$, we have $I_E = 6.493940$, $I_N =
2.404114$, while in the Wien limit, i.e., when $\xi\gg1$, we have $I_E=6$,
$I_N=2$.

Noting that $kT_{\rm r} \sim \widehat{E}/\widehat{n}$ and $e^{-\xi}
\sim \widehat{n}^4/C\widehat{E}^3$, a little trial and error gives
the following fitting function for the radiation
temperature $T_{\rm r}$ in terms of $\widehat{E}$ and $\widehat{n}$,
\begin{equation}
kT_{\rm r} = \frac{\widehat{E}/\widehat{n}}
{[3-2.449724 (\widehat{n}^4/C\widehat{E}^3)]}. \label{Tr2}
\end{equation}
The constants in the denominator have been chosen to give the correct
result in the Planck and Wien limits. There is a small deviation at
intermediate values of the chemical potential, but the fractional
error in the temperature is at worst no more than 0.04\%. In
the present photon-conserving Comptonization approach, instead of using
equation (\ref{Tr1}) which corresponds to blackbody Comptonization, 
we use equation (\ref{Tr2}) to calculate
$T_{\rm r}$.

Note that the ratio $\widehat{E}/\widehat{n}$ is equal to
  $2.7012\, kT_{\rm r}$ in the blackbody limit and equal to
  $3\,kT_{\rm r}$ in the extreme Wien limit. This is a small enough
  difference that, in light of our other approximations, we could
  simply use the blackbody value of the ratio for all regimes. The key
  point is that the radiation temperature $kT_{\rm r}$ should be
  estimated via the ratio $\widehat{E}/\widehat{n}$ (with some
  reasonable coefficient) if we wish to satisfy photon conservation,
  whereas any approach based purely on equation (\ref{Tr1}) will violate
  photon conservation. In this work, we use the fitting function
  (\ref{Tr2}) when we model photon-conserving
  Comptonizaiton.\footnote{Because of the negative sign in the
  denominator of equation~(\ref{Tr2}), the function can sometimes
  diverge and behave unphysically. This potential pathology (which has
  not been an issue in any of the simulations we have run so far,
  including those presented here) can be fixed by resetting the
    denominator to its minimum value of 2.7012 whenever it tries to go
    lower, or by choosing a different form of the fitting function
  (Jonathan McKinney, private communication).}

Since we have introduced a new variable $\widehat{n}$, we need an
evolution equation for this quantity. Here we use the fact that
absorption removes photons, emission adds photons, and scattering
leaves the number of photons unchanged. Thus, in the fluid frame, we
write the rate of change of photon number density as
\begin{equation}
\dot{\widehat{n}} = - \rho \left[
\left(\frac{\kappa_{\rm a}\widehat{E}/kT_{\rm r}}
{3-2.449724 (\widehat{n}^4/C\widehat{E}^3)}\right) - 
\left(\frac{4\pi\kappa_{\rm a}\widehat{B}/kT_{\rm g}}
{2.701178}\right) \right]. \label{nhatdot}
\end{equation}
The evolution equation for the photon number density $n$ is then
written as
\begin{equation}
\left(n u_r^\mu \right)_{;\mu} = \dot{\widehat{n}}, \label{nevol}
\end{equation}
where $u_{\rm r}^\mu$ is the four-velocity of the radiation frame, and
we have used the fact that $\dot{\widehat{n}}$ is a frame-invariant
quantity (because $\dot{n} = dN/dt \,dx_1dx_2dx_3$, and both the
numerator and the denominator are relativistic invariants).

In summary, under the photon-conserving Comptonization
approximation, we solve equations (\ref{Tmunu}), (\ref{Rmunu}) and
(\ref{nevol}), using equations  (\ref{Gi}), (\ref{CompG0}) and
(\ref{nhatdot}), where $\widehat{B}$ is given by equation (\ref{Tg})
and $T_{\rm r}$ is given by equation (\ref{Tr2}).

\section{Numerical methods}
\label{s.method}

To solve equations~\ref{Tmunu}-\ref{Gi}, we use the general relativistic radiation magnetohydrodynamical (GRRMHD) code
\texttt{KORAL} \citep{sadowski+koral,sadowski+koral2}, which employs a Godunov scheme
to evolve the conservation equations in
a fixed, arbitrary spacetime using finite-difference methods. 
The magnetic field is evolved via the induction equation, with
the divergence-free criterion being enforced using a flux-constrained \citep{toth-00}
scheme as described in \cite{gammie03}. The radiation field is evolved
through its energy density and flux, and the radiation stress-energy
tensor is closed by means of the M1 closure scheme \citep{levermore84}. For
details of the basic method see \cite{sadowski+koral2}. 

For the purpose of this work we adopt a simple prescription for
  the absorption opacity $\kappa_a$ \citep{katobook}, \be \kappa_{\rm
    a}=6.4\times 10^{22}\, \rho T_{\rm g}^{-7/2}\,{\rm cm^2g^{-1}}.
  \ee Note that the absorption opacity should in principle be
  a function of both gas and radiation temperatures. We ignore this
  complication here. Similarly, in the spirit of keeping things
    simple, we ignore the fact that the rate of change of photon
    number by absorption and emission is generally more complicated
    than the prescription given in equation (\ref{nhatdot}). These
    issues will be dealt with more accurately in future studies. For
  the scattering opacity we use 
\be \quad\kappa_{\rm es}=0.34 \,{\rm
    cm^2g^{-1}}. \ee

Thermal Comptonization within the ``blackbody'' approximation (Section~\ref{s.thermal}) has already been
included in \texttt{KORAL} and has been described in a previous paper \citep{sadowski+dynamo}. Implementing that scheme does not require
any modification to the basic algorithm in \texttt{KORAL}; we
merely need to modify the time
component of the fluid frame four-force, which describes the rate of
exchange of energy between gas and radiation, according to equation~(\ref{CompG0}).
The corresponding radiation four-force density $G^\mu$ then takes the form
\citep{sadowski+koral2},
\be
\label{e.gmu.thermal}
G_{\rm Compt}^\mu=\widehat G_{\rm Compt}^0 u^\mu,
\ee
where,
\begin{eqnarray}
\widehat{G}_{\rm Compt}^0 &=&
- \kappa_{\rm es}\rho\widehat{E} \left[\frac{4k(T_{\rm g}-T_{\rm
      r})}{m_{\rm g}c^2}\right]  \nonumber \\
&~~&\times\left[1+3.683 \left(\frac{kT_{\rm g}}{m_{\rm g}c^2}\right) 
+4 \left(\frac{kT_{\rm g}}{m_{\rm g}c^2}\right)^2\right]
\left[1+\left(\frac{kT_{\rm g}}{m_{\rm g}c^2}\right)\right]^{-1},
\label{CompG02}
\end{eqnarray} 

The photon-conserving Comptonization scheme introduced in this work
(Section~\ref{s.ncompton}), on the other hand, requires a modified
algorithm. One more conserved quantity -- the photon
number density $n$ -- has to be evolved, for which we use equation~(\ref{nevol}).
The source term on the right hand side of (\ref{nevol}) describes the
change of photon number due to emission and absorption by the
gas, and is proportional to the absorption opacity $\kappa_{\rm a}$.
As in the case of other radiative source terms
(equations.~\ref{G00}-\ref{Gi}), this term may become stiff in an optically thick
medium. Therefore, it is necessary to apply it in an implicit way. For
this purpose we modify the original semi-implicit operator \citep{sadowski+koral}, which applies
the standard radiative source terms, to include the evolution of the
photon number density. The set of equations solved in the implicit
step is now,
\begin{eqnarray}
\hspace{2cm}(\rho u^t)_{(i+1)}-(\rho u^t)_{(i)} &=&0, \label{imp-1} \\
T_{\nu,(i+1)}^t-T^t_{\nu,(i)} &=& \Delta t G_{\nu,i+1}, \label{imp0} \\
R_{\nu,(i+1)}^t-R^t_{\nu,(i)} &=& -\Delta t G_{\nu,i+1}, \label{imp1}\\
(n u_{\rm r}^t)_{(i+1)} - (n u_{\rm r}^t)_{(i)}&=&\dot{\widehat n}_{(i+1)}, \label{imp2}
\end{eqnarray}
where subscripts $(i)$ and $(i+1)$ indicate values at the beginning
and end of the current
time step, respectively. Due to the symmetry of the problem, it is
enough to iterate just one set of primitives, either radiative or
hydrodynamical, extended by the photon number density, and to use the
conservation of energy and momentum to recover the remaining
quantities.

The new system of equations is solved for values at the new moment of
time (denoted by $(i+1)$) using the Newton-Raphson method in five
dimensions. The computational cost
increases because of the extra dimension by a factor of $\lesssim
2$ in
most cases.

\section{Simulations}
\label{s.sims}

\subsection{Initial setup}

To assess how important the proper treatment of Comptonization is for
accretion disks we carried out a set of
simulations of super-critical accretion flows on a $10 M_\odot$ BH. In order to obtain a relatively
large radial range of inflow/outflow equilibrium, we decided to perfom the
simulations in 2D assuming axisymmetry. The poloidal magnetic field was maintained 
by means of the
mean-field magnetic dynamo model of \cite{sadowski+dynamo}.

We
performed three simulations, each initiated in exactly the same way as
described in  \cite{sadowski+koral2}, with parameters given
in Table~\ref{t.models}. The only difference between the runs was
the way Comptonization was treated: (i)
basic ``blackbody Comptonization'' prescription (Section~\ref{s.thermal}), (ii)
improved ``photon-conserving Comptonization'' (Section~\ref{s.ncompton}), (iii)
no Comptonization at all. For the run with photon-conserving
Comptonization
we had to specify the initial photon density. Because the radiation
and gas in the initial torus satisfy local thermal equilibrium,
we set the radiation and gas temperatures to be equal and
set the photon number density according to the blackbody formula,
\be
\widehat n = \frac{a_{\rm rad} T^3}{2.7012 k_{\rm B}}.
\label{e.nlte}
\ee
Hereafter, we use the
gravitational radius $R_{\rm g}=GM/c^2$ as the unit of length, and $R_g/c$
as the unit of time.

\begin{table}
\centering
\caption{Model parameters}
\label{t.models}

\begin{tabular}{lr}
\hline
\hline
$M_{\rm BH}$ & $10 M_\odot$ \\
$a_*$ &   0.0  \\
$\rho_{\rm max}$ & $4.3\times 10^{-3}\,\rm g/cm^3$ \\
$\beta_{\rm max}$  &  10.0 \\
$N_R$ x $N_\theta$  &   252 x 234\\
$R_{\rm min}$ / $R_{\rm max}$ / $R_0$ / $H_0$  &   1.85 / 1000 / 0 / 0.6\\
$t_{\rm max}$ &  48,000\\
$\langle \dot M \rangle /\Medd$  &   10.2 (blackbody Compt.) \\
& 11.2 (N-Compt.) \\
& 9.2 (no Compt.) \\
\hline
\hline
\multicolumn{2}{l}{$\rho_{\rm max}$ - maximal density of the initial torus }\\
\multicolumn{2}{l}{ $\beta_{\rm max}$ - maximal value
of initial total to magnetic pressure ratio}\\
 \multicolumn{2}{l}{$t_{\rm max}$ - duration of simulation }
\end{tabular}
\end{table}

\subsection{Accretion flow properties}
\label{s.results}

\begin{figure}
\includegraphics[width=.99\columnwidth]{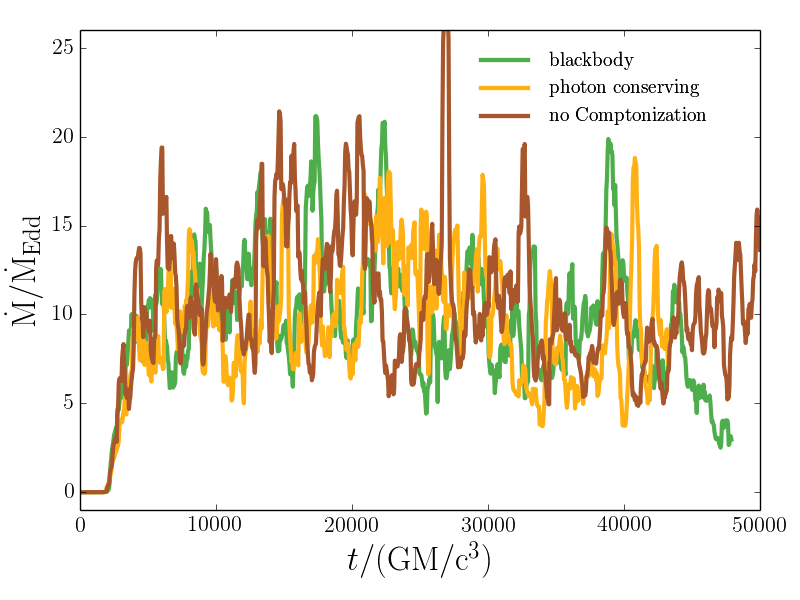}

\caption{Mass accretion rate at the BH as a function of time for the three
  models considered.}
\label{f.mdots}
\end{figure}

All three simulations were run up to a final time of $t_{\rm max}=48,000
 $. The region of inflow/outflow equilibrium extended up to a
 radius $R\approx 30$ in the equatorial plane, and to a much larger
 distance in the funnel region near the axis (where the radial
 velocity of the outflowing gas is much higher, $\gtrsim 0.1c$). 
The mean accretion rate for all the models was
$\sim 10\Medd$ (Table~\ref{t.models}). The time history of the mass accretion rate through
the BH horizon is shown in Fig.~\ref{f.mdots}. Although the profiles
are highly variable, as expected for a turbulent accretion flow,
there are no significant differences in the three profiles.
This suggests that the precise treatment of
Comptonization has little impact on the the gross dynamical properties
of the gas.

\begin{figure*}
blackbody \hspace{4cm} photon-conserving \hspace{3cm} no Comptonization\\
\includegraphics[width=.38\textwidth]{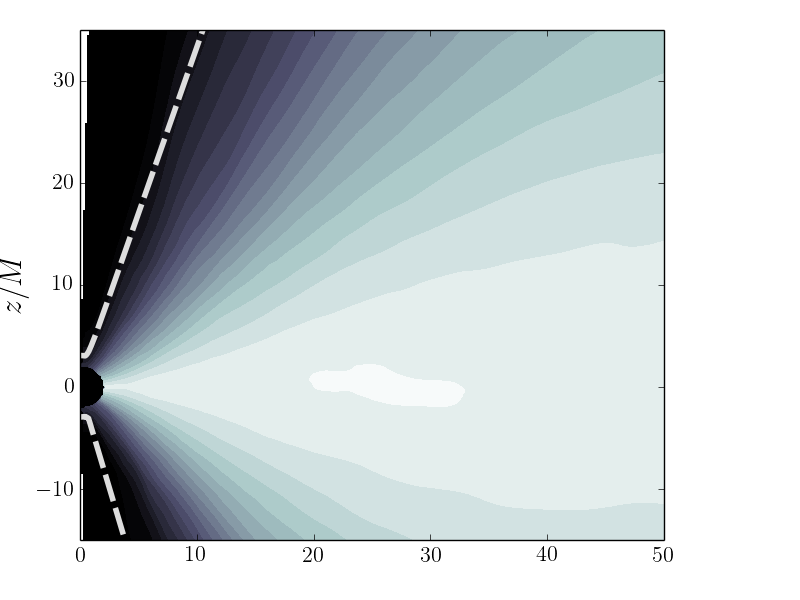}\hspace{-1.4cm}
\includegraphics[width=.38\textwidth]{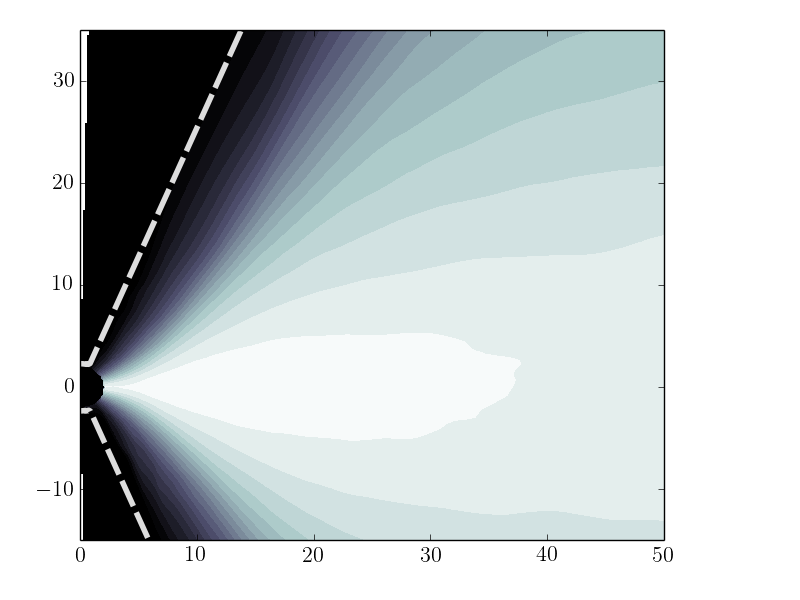}\hspace{-1.4cm}
\includegraphics[width=.38\textwidth]{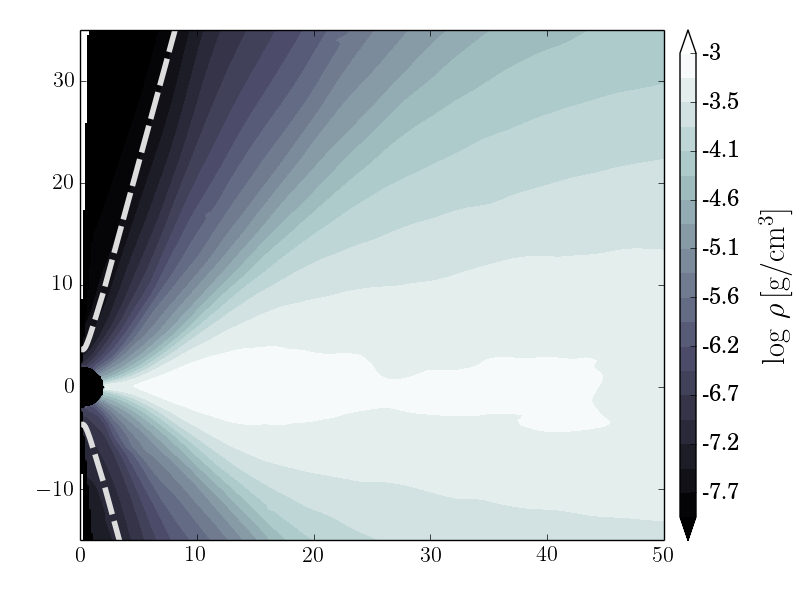}\hspace{-0cm}
\includegraphics[width=.38\textwidth]{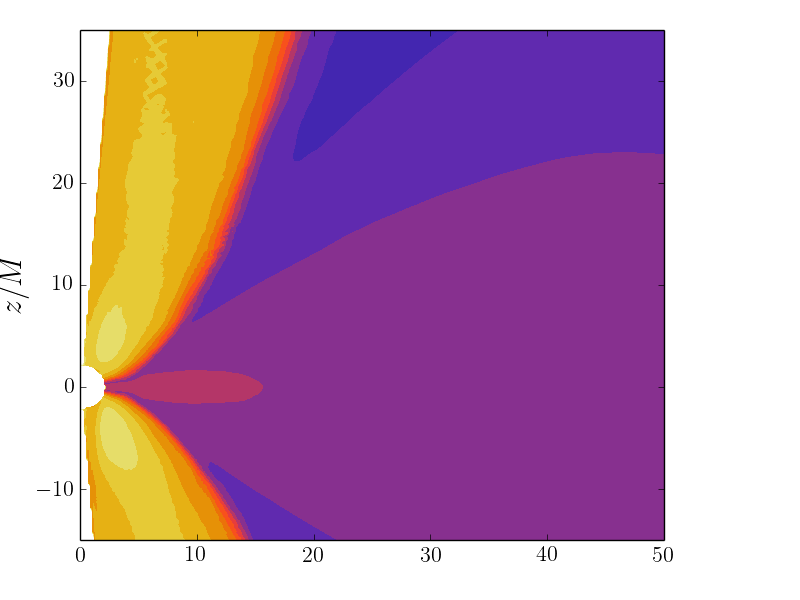}\hspace{-1.4cm}
\includegraphics[width=.38\textwidth]{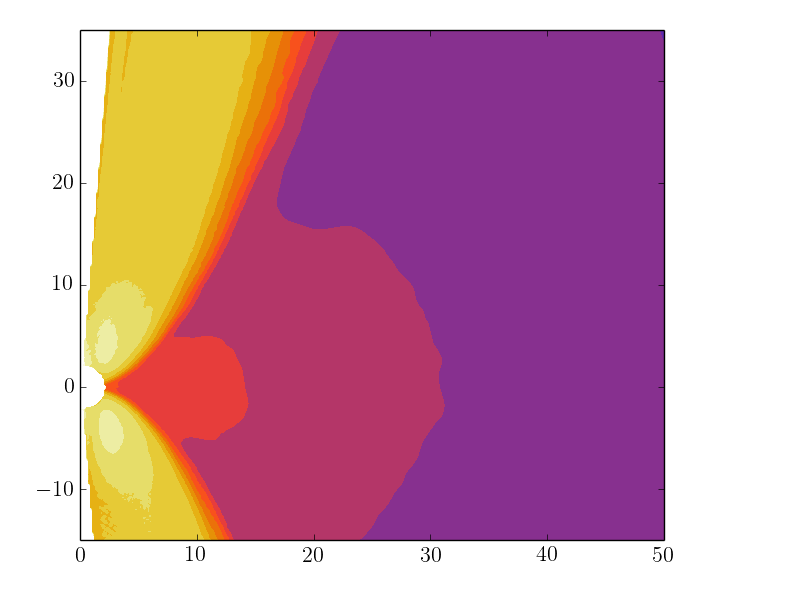}\hspace{-1.4cm}
\includegraphics[width=.38\textwidth]{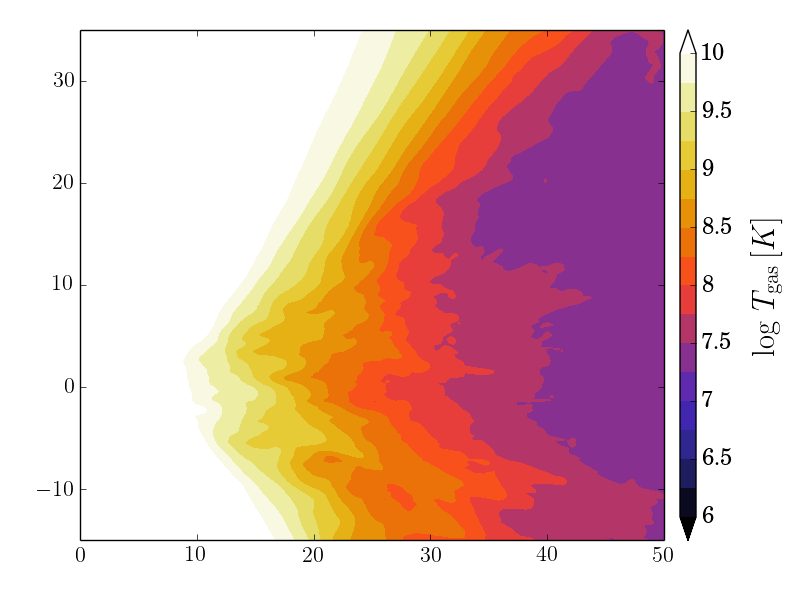}\hspace{-0cm}
\includegraphics[width=.38\textwidth]{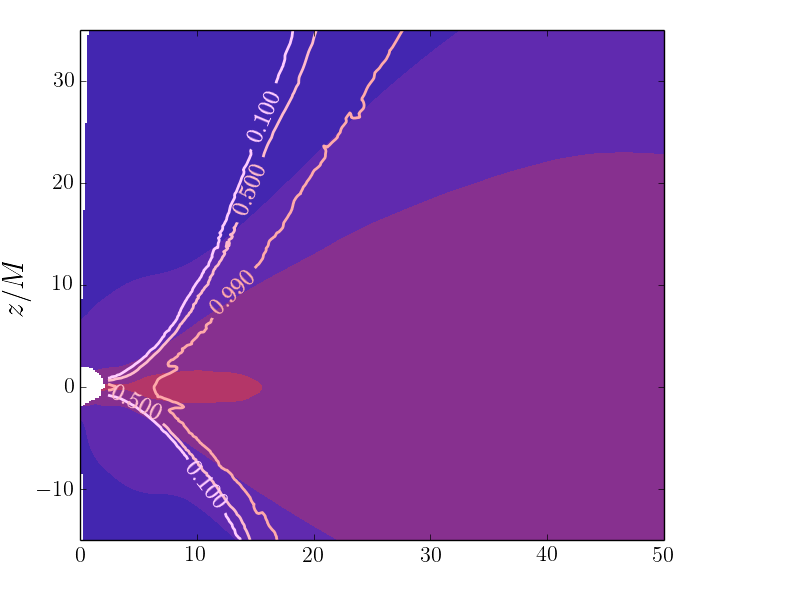}\hspace{-1.4cm}
\includegraphics[width=.38\textwidth]{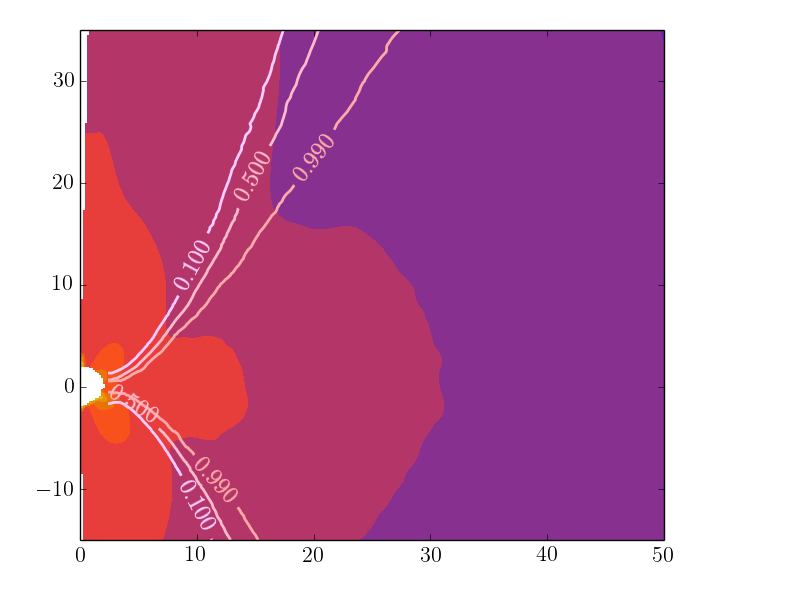}\hspace{-1.4cm}
\includegraphics[width=.38\textwidth]{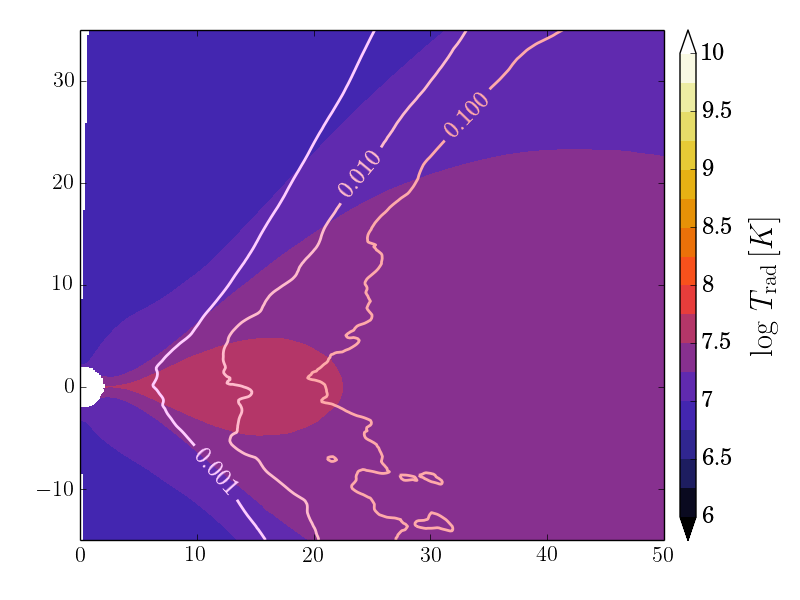}\hspace{-0cm}
\includegraphics[width=.38\textwidth]{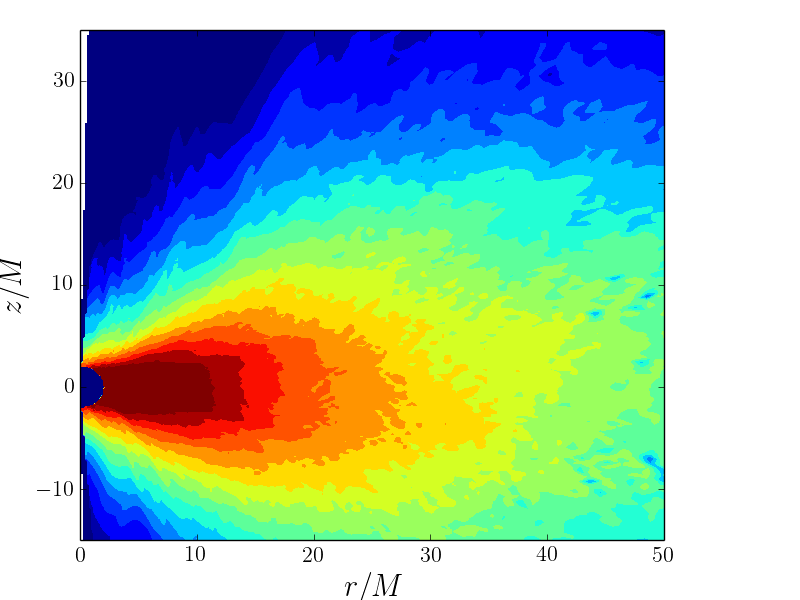}\hspace{-1.4cm}
\includegraphics[width=.38\textwidth]{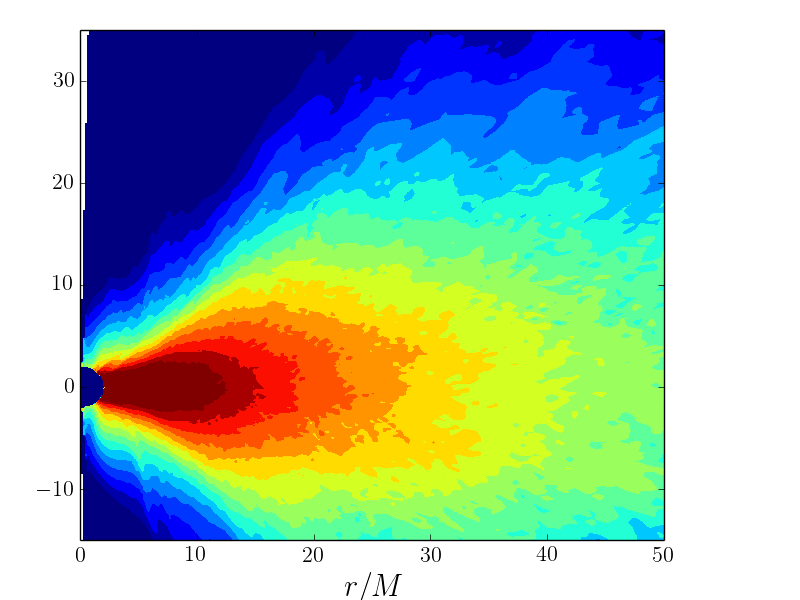}\hspace{-1.4cm}
\includegraphics[width=.38\textwidth]{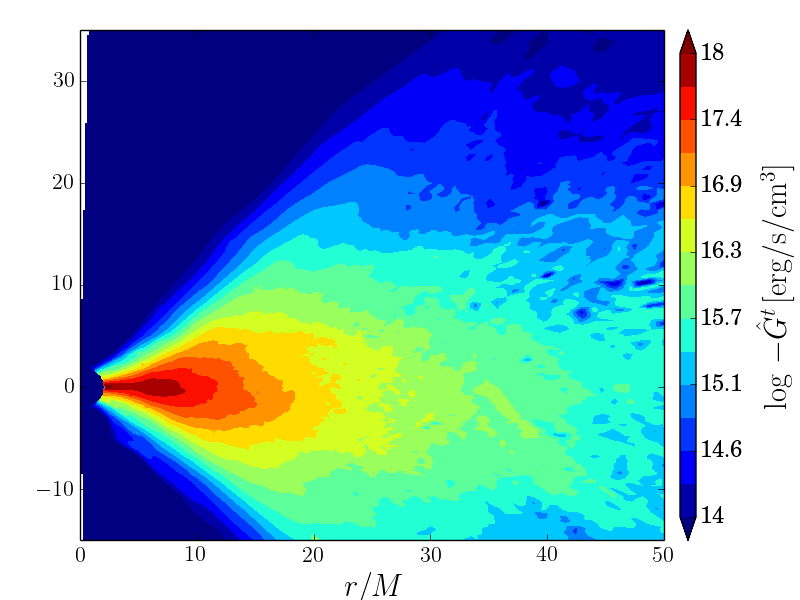}\hspace{-0cm}
\caption{Distributions of density (top row), gas temperature (second row)
  radiation temperature (third row), and 
  cooling rate (bottommost row) for models with blackbody Comptonization
  (left), photon-conserving Comptonization (middle), and no
  Comptonization (right panels). Plots correspond to data
  averaged over $t=24000\div48000$. The dashed lines in the
  density plots reflect the location of the scattering photosphere as
measured from infinity along fixed polar angle. The contours on the
radiation temperature plots (third row) reflect the
radiation to gas temperature ratio.}
\label{f.9plots}
\end{figure*}

A similar conclusion is reached when comparing the time- and $\phi$-averaged distributions
of density in the poloidal plane, shown in the top row of panels in
Fig.~\ref{f.9plots}. From the left, the three panels correspond to
blackbody Comptonization, photon-conserving
Comptonization, and no Comptonization, respectively. 
In all three simulations, the accretion flow is geometrically thick,
with an average density scale height $H/R\sim0.3$. The
densities at the equatorial plane are similar in all three runs, with $\rho \sim10^{-3}
~{\rm g\,s^{-1}}$. The dashed lines in the panels show the location of
the scattering photospheres as measured from infinity along fixed
polar angle $\theta$. In all models the polar region is optically thin
(i.e., the observer looking into the funnel can directly see
radiation coming from the vicinity of the BH horizon). The size of
this region is largest for the
photon-conserving model, but the difference is not large.

Although the densities in the three runs are similar, the gas
and radiation temperatures, shown respectively in the second and third rows of panels in
Fig.~\ref{f.9plots}, are significantly different. In scattering
dominated media, and in stellar mass BH accretion disks in
particular, Comptonization often dominates the energy transfer rate
between the gas and radiation; without Comptonization, the coupling is
very weak. 

Turbulent dissipation in the disk heats the gas, the rate of heating
depending mostly on the properties and evolution of the MHD turbulence,
which is insensitive to the way we treat Comptonization. 
When the
gas subsequently cools, the dissipated heat energy is transfered to radiation.
To
sustain a given rate of energy transfer from gas to radiation, one can either have efficient
coupling, with gas and radiation temperatures close to each other, or
weak coupling, with a large difference in the two temperatures.
In the case of the simulation with no
Comptonization (right panels), the coupling is weak, therefore the gas
temperature becomes very large. This effect is significant in the inner
regions of the disk and is particularly evident in the coronal
and polar regions, where the gas temperature exceeds $10^{10}\,\rm K$, while
the radiation temperature stays around $10^7\,\rm K$. 
In contrast, for the two models that allow Compton coupling, viz.,
blackbody Comptonization and photon-conserving Comptonization,
the gas temperature is much lower, and is also closer to the radiation
temperature, reflecting the fact that the coupling between the
gas and the radiation field is strong.  As in the case of the
no Comptonization model, the gas
temperature tends to be higher than the radiation temperature here as well,
but the difference, even in the coronal region, is no longer very extreme. 

\begin{figure}
\includegraphics[width=1.0\columnwidth]{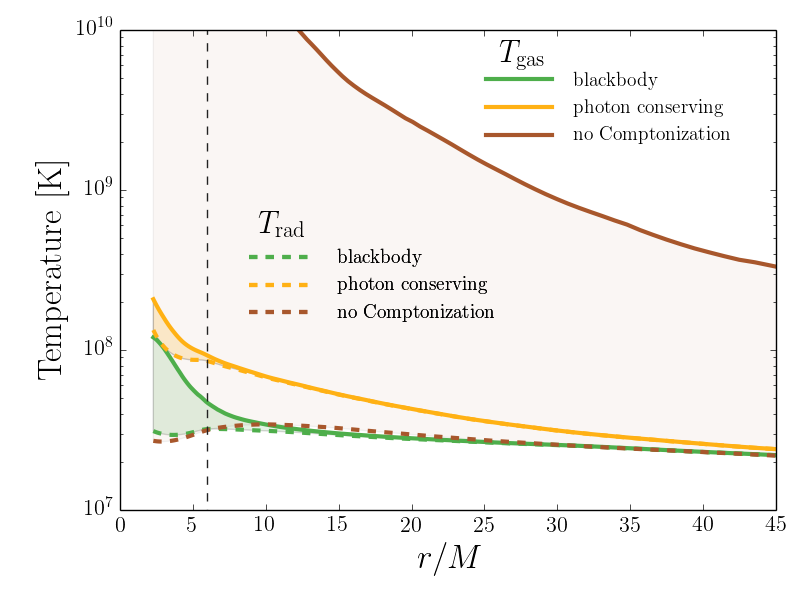}

\caption{Radial profiles of density weighted gas (solid) and radiation
  (dashed lines) temperatures in the three models. The shaded regions reflect the
  difference between the two temperatures.}
\label{f.temps}
\end{figure}

Radial profiles of the gas and radiation temperatures (these are
density-weighted, and therefore are dominated by the gas in the disk
interior rather than the corona) are shown in Fig.~\ref{f.temps}. The
average gas temperature at radius $R=15$ is approximately
$5\times10^{9}$, $6\times10^{7}$, and $3\times10^{7}\,\rm K$ for
models with no, photon-conserving, and blackbody Comptonization,
respectively. The overheating of the gas in the no Comptonization
model is dramatically evident.  The other two models (which include
Comptonization) have lower gas and radiation temperatures and the two
temperatures are nearly equal in the bulk of the disk ($R \gtrsim 10$). The actual
temperatures are, however, not the same for the two models.  The
photon-conserving model is hotter by a factor of 2 (measured by
either gas or radiation temperature). This is a significant difference
and will have a noticeable effect on the spectral properties of the
escaping radiation.

\begin{figure}
\includegraphics[width=1.15\columnwidth]{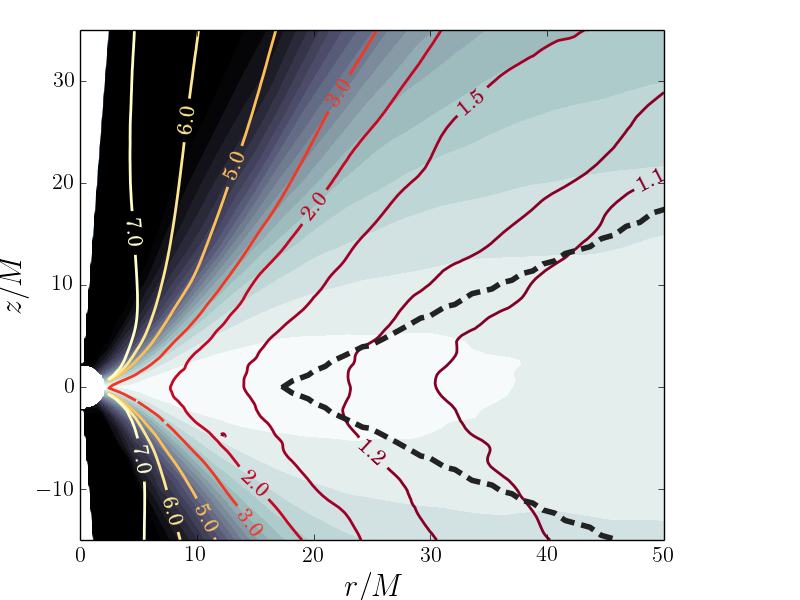}
\caption{Color correction factor $f_{\rm col}$ (equation~\ref{eq.fcol})
  corresponding to the photon-conserving Comptonization
  model. The colors reflect the distribution of gas density and the
  black dashed line shows the location of the effective photosphere
(equation~\ref{effphoto}).}
\label{f.fcol}
\end{figure}

The blackbody Comptonization and no Comptonization models assume
that the radiation spectrum is blackbody. Therefore, in these models
the radiation temperature $T_{\rm r}$ is a
function only of the radiative energy density in the fluid frame,
\be
\label{Tr3}
T_{\rm r}=\left(\frac{\widehat E}{a}\right)^{1/4}\equiv T_{\rm r,\,BB}.
\ee
For the photon-conserving Comptonization model, however, we track the
number density of photons and use equation~(\ref{Tr2}) to compute $T_{\rm r}$. This
temperature is equal to $T_{\rm r,\,BB}$  only in those regions
of the flow where the radiation happens to be black
body, e.g., deep inside the disk. In other regions, the local number density of photons
deviates from the blackbody value, and so do the temperature and spectrum. We define the
color correction factor as the ratio of the real radiation temperature to the blackbody temperature,
\be
\label{eq.fcol}
f_{\rm col}=\frac{T_{\rm r}}{T_{\rm r,\,BB}}.
\ee
Values of $f_{\rm col}$ larger than unity correspond to a radiation
spectrum that is harder than blackbody. Under the gray approximation we have adopted
in this work, we do not have detailed information on the spectral
shape of the radiation field. Therefore, spectral distortion from blackbody is described by a single
number $f_{\rm col}$.  More detailed, frequency-dependent radiative transfer is
required if we wish to obtain additional information.

By definition, the color correction factor (equation~\ref{eq.fcol}) is equal to
unity for the blackbody and no-Comptonization models, since the radiation
spectrum is assumed to be blackbody. In
the photon-conserving Comptonization model, however, it can (and does) deviate from unity.
The contours in Fig.~\ref{f.fcol} show the color correction
factor for this model plotted on top of the gas density
distribution. The black dashed line shows the location of
the effective photosphere, estimated by integrating the effective
optical depth from the axis along $\theta$ at fixed \textit{radius} and
identifying the angle at which it equals $2/3$:
\be
\int_0^\theta \rho \sqrt{\kappa_{\rm a}(\kappa_{\rm a}+\kappa_{\rm
    es})}\,\sqrt{g_{\theta\theta}}\,d\theta' = 2/3.
\label{effphoto}
\ee

Deep in the disk and at large radii ($R\gtrsim 30$), where the
disk is opticaly thick to both scattering and absorption, $f_{\rm col}$
is close to unity and  the radiation spectrum is
close to blackbody. The color correction increases (the spectrum becomes
harder) as the radiation gets closer to the disk surface  --- the
photons are up-scattered by hot gas and gain energy. Near the edge of
the funnel, the color correction factor reaches $f_{\rm col}\approx
5.0$, which reflects how strongly scattering modifies the radiation
field.\footnote{It is interesting to note that similar values of
    the color correction factor were obtained by \cite{kawaguchi-03}
    who estimated Comptonized spectra of slim accretion disks.}
This large value suggests that the spectrum would be strongly dominated by
a Compton hump, and the original black body component would be
insignificant \citep[compare][]{kawashima+12}.

The rate of energy transfer from gas to radiation is the cooling rate of the gas and
describes how effectively the heat energy released by turbulent dissipation
goes into the radiation
field. The bottommost panels of Fig.~\ref{f.9plots} show the
distribution of the time-averaged cooling rate for the three
models. For all three simulations, the highest rates of energy transfer are at the equatorial
plane and in the innermost
region ($R\lesssim 15$). However, the vertical extent of this
region as well as the magnitude of the cooling differ. The no Comptonization
model shows the least amount of cooling both in the funnel region near the axis and at the equatorial plane. For the thermal and photon-conserving
Comptonization models, the cooling region extends
further towards the axis and the magnitude of the cooling rate is also 
larger.



\begin{figure}
\includegraphics[width=1.0\columnwidth]{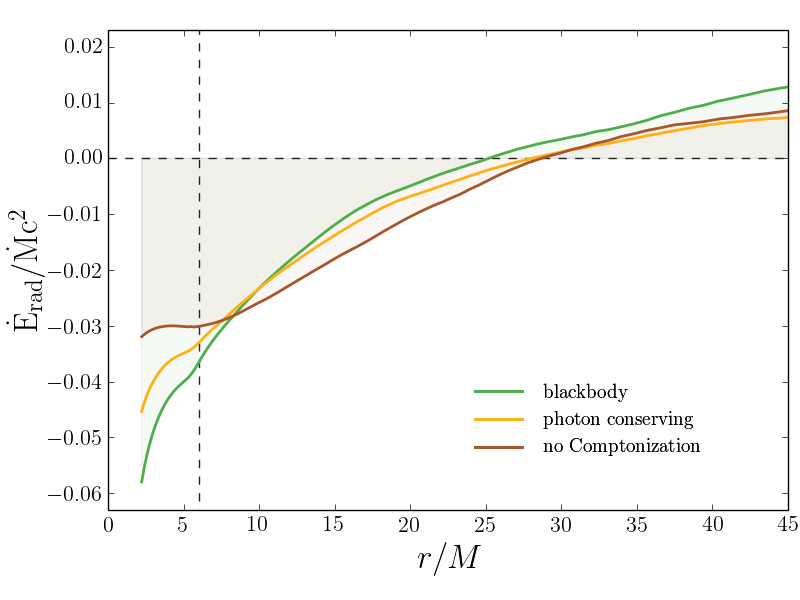}
\caption{Radiative luminosity $\dot E_{\rm rad}$ (equation~\ref{Edotrad}) as a function of radius for the three models.}
\label{f.radflux}
\end{figure}

Because accretion flows are quasi-stationary, the
energy transfered from gas to radiation via cooling
must be removed in some fashion. In optically thick disks, this can
happen in two ways --- either by radiation flowing out of the disk or
by radiation being trapped in the gas and carried into the BH with the flow 
(advected). 
Fig.~\ref{f.radflux}
is a plot of the radiative luminosity as a function of radius for the
three models, calculated as follows,
\be
\dot E_{\rm rad}=\int_2^R R^r_t \sqrt{-g}\, {\rm d}A,
\label{Edotrad}
\ee
where $\sqrt{-g}$ is the metric determinant and ${\rm d}A = 2\pi r^2 d\theta$ is a
differential surface element at radius $r$.
Negative values of this quantity
correspond to regions where radiation advection into the
BH dominates. The radius where $\dot E_{\rm rad} = 0$
reflects the location of the effective
trapping radius of the accretion flow. 

All three runs show significant advection of photons.
Only outside
radius $R\approx 25$ does the net flux of radiation point outward. To estimate the total amount of energy put into the
radiation field we take the luminosity in inflowing photons at the BH
and add it to 
the luminosity in outflowing photons at
radius $R=45$. The total  energy release in
radiation is roughly $7\%$, $6\%$, and $4\%\dot M c^2$ for the
blackbody, photon-conserving, and no Comptonization models,
respectively\footnote{We stress here that these numbers do not
  correspond to the standard definition of radiative efficiency,
  which includes only radiation escaping to a distant observer.}. These numbers are in agreement with the discussion of
the cooling rate in the previous paragraphs, viz., not including
Comptonization underestimates the cooling rate.

\begin{figure}
\includegraphics[width=1.0\columnwidth]{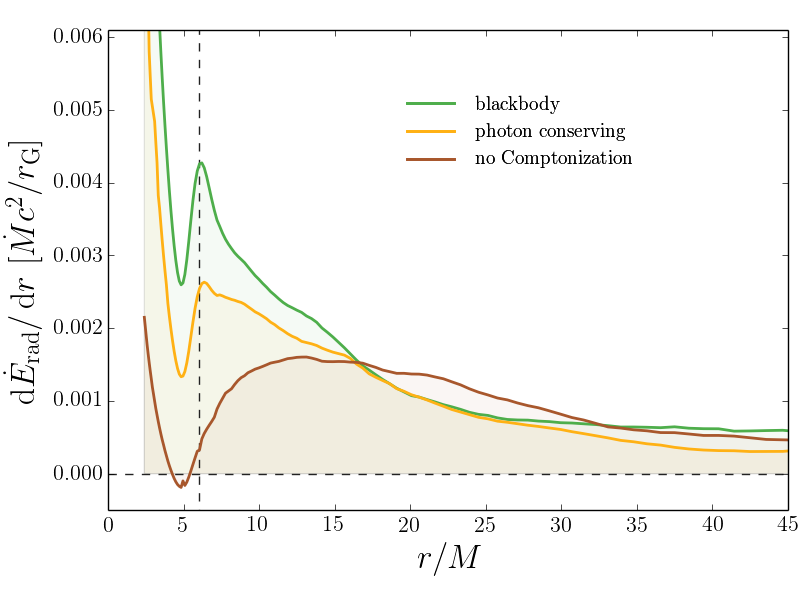}
\caption{Profiles of radiative emission per unit radius for the three models.}
\label{f.dissrate}
\end{figure}

By taking the radial derivative of the radiation luminosity $\dot E_{\rm rad}$, we obtain
the local photon generation rate or the gas cooling rate. This quantity is shown
in Fig.~\ref{f.dissrate}. The emission profiles more or less overlap in the outer
($R>15$) region but are significantly different inside that radius.
The maximum amount of energy is put into radiation in the blackbody Comptonization model, and the least
when Comptonization is not taken into
account. In the latter model, there is hardly any emission inside the
marginally stable orbit.

\section{Discussion}
\label{s.discussion}

In this work we have introduced an improved method for treating
Comptonization in computational fluid dynamics. The suggested approach
conserves the number of photons and allows for a non-blackbody
spectrum of radiation. It is fully covariant and may be applied to
simulations performed in general relativity. However, the method is
still very approximate, since it assumes a specific spectral shape for
the radiation, viz., a Bose-Einstein distribution function
(eq.~\ref{BE}). Also, the proposed algorithm is valid only for gray
radiative transfer.  Nevertheless, we believe this ``photon-conserving
Comptonization'' method is a significant improvement over the traditional
``blackbody Comptonization method''.

To assess the effect of various treatments of Comptonization in
modeling accretion disks, we performed a set of three simulations
corresponding to a super-critical disk accreting at $10\dot M_{\rm
  Edd}$. We implemented the new photon-conserving approach (described
in detail in Section~\ref{s.ncompton}) as well as
blackbody Comptonization (Section~\ref{s.thermal}) and considered also
a model with no Comptonization. 
We find that the
density distribution and the accretion rate are insensitive to the way we
model Comptonization, whereas the gas temperature and the properties of
the radiation field depend strongly on which version of Comptonization we use.
In particular, simulations with no Comptonization produce
unphysically high gas temperatures. This is because of inefficient coupling between
gas and radiation. The blackbody approach, on the other
hand, noticeably underestimates the gas and radiation temperatures. In
making these statements, we are working on the assumption that the
photon-conserving model, being the most sophisticated of the three, is
closest to the true solution.

By comparing the three simulations we find that the impact
  of Comptonization is not limited to the coronal regions (as
  claimed by some authors), but that it actually affects the gas
  properties throughout the disk, even deep inside the effective
  photosphere. With no Comptonization at all, the coupling between gas
  and radiation comes only from the absorption opacity, which requires
  significantly different gas and radiation temperatures. With
  Comptonization, the optically thick gas in the bulk of the disk is
  at lower temperatures, closer to the radiation temperature.

The photon-conserving model provides one additional piece of information
compared to the other two models, viz., the color correction factor $f_{\rm col}$
as a function of position. This factor, which measures the ratio of the radiation
temperature to an effective blackbody temperature (eq.~\ref{eq.fcol}), describes
the amount of spectral hardening caused by scattering and Comptonization.
We find that $f_{\rm col}$ can deviate substantially from unity, and can be
as large as 5 or 6 at the photosphere inside the funnel. Such
  high color correction factors will characterize radiation escaping
  through the polar funnel region, and will be visible to observers viewing at
  small polar angles. For larger inclinations, most of the observed radiation would
  come from the disk photosphere located far from the BH (this region is not
  resolved in our simulations) and would likely not show such strong Comptonization.

Another quantity that is significantly affected by the version of
Comptonization that one uses is the amount of cooling, which
determines the ultimate luminosity (both in advected and escaping
photons). The blackbody and no-Comptonization models over- and
under-estimate the cooling rate, respectively.

Taking into account the modest computational cost of the
photon-conserving Comptonization approach, we recommend using it in
simulations of accretion disks. It is true that blackbody
Comptonization reproduces reasonably well the dynamical properties of
the accretion flow and the cooling rate. However, the gas and
radiation temperatures obtained via this approach are significantly
under-estimated.  Therefore, this model cannot be trusted for
applications in which the temperature is of interest, e.g., spectral
modeling.

To complete this study we need to assess how the treatment of
Comptonization affects the spectrum as seen by a distant observer. 
This requires post-processing with a multi-dimensional radiative
transfer solver that works in general relativity and allows for
frequency-dependent opacity \citep[e.g.,][]{zhu+15} and
Comptonization. 

\section{Acknowledgements}

The authors thank Jonathan McKinney for helpful discussions and comments. AS acknowledges support
for this work 
by NASA through Einstein Postdoctotral Fellowship number PF4-150126
awarded by the Chandra X-ray Center, which is operated by the
Smithsonian
Astrophysical Observatory for NASA under contract NAS8-03060. AS thanks
Harvard-Smithsonian Center for Astrophysics for its hospitality.
RN was
supported in part by NSF grant AST1312651 and NASA grant TCAN
NNX14AB47G.
The authors acknowledge computational support from NSF via XSEDE resources
(grant TG-AST080026N), and
from NASA via the High-End Computing (HEC) Program
through the NASA Advanced Supercomputing (NAS) Division at Ames
Research Center.
 
\bibliographystyle{mn2e}

\end{document}